\documentstyle[epsfig,11pt]{article}

\def\beq{\begin{equation}}
\def\eeq{\end{equation}}

\title{\bf Scattering amplitude and pomeron loops in 
perturbative QCD at large  $N_c$}
\author{M.A.Braun, A.N.Tarasov  \\
St.Petersburg State University, Russia}

\begin{document}

\maketitle

{\bf Abstract}

The  amplitude for the collision of two hadrons with the lowest order
pomeron loop is calculated. Numerical calculations show that the loop
contribution to the amplitude begins to dominate the single pomeron
exchange at rapidities $8-10$. Full dependence of the triple-pomeron
vertex on intermediate conformal weights is taken into account.

\section{Introduction}
In the framework of QCD with a large number of colours $N_c\gg1$,
strong  interactions are mediated by the exchange of
interacting BFKL pomerons, which split and fuse by triple pomeron  vertices.
This picture can be conveniently described by an effective non-local
quantum field theory ~\cite{braun1}. From the BFKL approach this theory
inherits  the fundamental property of
conformal invariance, which is only broken by interactions with
colliding hadrons. In terms of the relevant Feynman diagrams
contributions standardly separate into tree diagrams and diagrams with
pomeron loops, which corresponds to division into quasi-classical
contribution and quantum corrections. The relative contribution of loops
is characterized by the parameter $\alpha_s^4N_c^2\exp(\Delta_{BFKL}y)$,
where $y$ is the rapidity and $\Delta_{BFKL}$ is the intercept of the
BFKL pomeron (minus unity)~\cite{mueller}. So although the loop contributions are
damped by factor $1/N_c^2$ as compared with the tree contributions,
which depend on $\alpha_sN_c$, the large exponential factor makes
their relative order grow with energy. It is of vital importance then
to estimate the role of pomeron loops at present energies to have some
idea of the validity of the quasi-classical methods which are
currently applied to the analysis of the high-energy scattering in the
framework of QCD,  such as the Balitski-Kovchegov equation ~\cite{bal,kov},
its generalization to nucleus-nucleus collisions ~\cite{MB06} and the
Color Glass Condensate approach (see the review ~\cite{CGC} and
references therein).

In principle calculation of the loop contribution to the BFKL exchange
in the lowest order seems to be straightforward. One has all the instruments
for this goal: the BFKL propagator and the triple pomeron vertex.
The main and only obstacle is the most complicated form of the latter,
which converts a realistic calculation of the loop contributions
into a formidable task. For these reasons there were quite a few attempts in
this direction, all suffering from crude approximations and uncertainties.
In ~\cite{pechan,baryva} the pomeron loop was calculated under the
assumption that the triple-pomeron vertex is independent of
conformal weights. Apart from that, simplified (asymptotic)
expressions were used for the BFKL propagator into which the loop is
inserted, in spite of the fact that with the loop growing much faster this
is not allowed. The results in ~\cite{pechan,baryva} do not agree (probably
due to differences in numerical coefficients). In any case in ~\cite{baryva}
it was found that the magnitude of the loop is so small
that it gives no significant contribution up to extraordinary high energies
(rapidities of the order of 40). Later in ~\cite{MB09} a step forward
was taken to take into account the dependence of the loop on inner
conformal weights. This allowed to study the relative magnitude of
the loop for the BFKL propagator with a fixed (dominant) conformal weight.
This relative magnitude was found to be much larger: the loop begins to
dominate already at rapidities of the order of 10$-$15. Still this result did
not solve the real physical problem, the contribution of the loop to the
scattering amplitude, which is obtained after integration over all
conformal weights and requires knowledge of the triple pomeron vertex
as a function of three conformal weights.

Actually the formal expression for the triple pomeron vertex as a function
of three conformal weights was derived by G.Korchemsky long ago
~\cite{KC99}. Nevertheless its
complexity has till now not been used for practical calculations.
In this paper we try to exploit Korchemsky's formulas to actually
calculate the contribution of the pomeron loop to the hadronic scattering
amplitude in full rigour. This contribution turns out to be quite large and
becomes comparable to the tree result at rapidities of the order of 10.

We stress that in this paper we limit ourselves to the loop contribution in the
lowest order (next-to-leading order in $1/N_c^2$). At high energies it grows
faster than the single pomeron exchange contribution.
Since higher order loops grow still faster, in the high-energy limit one has to 
sum all the loops to know the behaviour of the amplitude.
 We do not pretend to solve this problem here.
Our aim is only to find the threshold energy above which loop
contributions have to be taken into account. 

Our calculations are done within the BFKL theory with a fixed coupling constant.
Inclusion of the running coupling may significantly change our results.
However, in absence of conformal invariance,
this requires extraordinary computational efforts and is postponed for
future studies. Also we limit ourselves with the simplest case of proton-proton
scattering. Proton-nucleus and especially nucleus-nucleus scattering require
special attention and will also be considered in future publications. We expect
that the single loop influence for this processes will start at the same energy as for
the proton-proton case.

We have to note that in literature one can find several attempts to sum 
all loop contributions 
in the colour dipole approach or in the so-called reaction-diffusion formulation
of the scattering
mechanism ~\cite{lo1,lo2,lo3,lo4,lo5}. In particular in ~\cite{lo1,lo2,lo3}
an evolution equation for the gluodynamics was written, which in principle
allows to include pomeron loops to all orders. However the complexity of
the equation makes one recur to some drastic approximations to obtain
some concrete results, like considering simplified reaction-diffusion models.
In ~\cite{lo4} a similar Langevin equation was derived from the interacting pomeron
theory of \cite{MB06}. However the solution of this equation was not attempted.
In ~\cite{lo5}  pomeron loops were approximately imitated
by multiple pomeron exchanges with "renormalized" vertices. However uncontrollable
approximations made in the course of the derivation render it rather dubious.  
Thus the reliability of all these results remains open.

The paper is organized as follows. We recall some basic elements of the
conformal approach in the next section. In section 3 we use this results
to calculate the single-loop contribution to the scattering amplitude with more
or less arbitrary form of the impact factors. The last section presents our
numerical results and discusses influence of the loop contribution on the
scattering amplitude.

\section{Conformal invariant technique}

The conformal technique is based on Moebius (conformal) invariance of the
BFKL approach. It uses expansion of the pomeron propagator and
triple-pomeron vertex in orthogonal functions of the conformal basis.
The Mobius invariance drastically simplifies this expansion.
One can use these results and rewrite expression for any Feynman diagram in
terms of conformal basis. In this case integrations over gluon coordinates
transform  into summations over conformal weights.

The BFKL pomeron propagator satisfies the BFKL evolution equation:
\beq
(\frac{\partial}{\partial y}+H_{BFKL})g_{y-\bar{y}}(r_1,r_2;\bar{r}_1,\bar{r}_2)=
\delta(y-\bar{y})\nabla_1^{-2}\nabla_2^{-2}\delta^{(2)}(r_1-\bar{r}_1)
\delta^{(2)}(r_2-\bar{r}_2),
\eeq
where $y$ is the rapidity,
\beq
H_{BFKL}=\frac{\bar{\alpha}}{2}\Bigl(\ln p_1^2+\ln p_2^2+\frac{1}{p_1^2}
\ln r^2_{12}p_1^2+\frac{1}{p_2^2}\ln r^2_{12}p_2^2-4\psi(1)\Bigr)
\eeq
is the BFKL Hamiltonian with $\bar{\alpha}=\alpha_s N_c/\pi$ and
$p_i$ and $r_i$ are  transverse momenta and coordinates of the
reggeized gluons,
$r_{ij}=r_i-r_j$.

The solution of this equation was found in terms of expansion in the conformal
basis \cite{LL86}:
\beq
g_{y-y'}(r_1,r_2;\bar{r}_1,\bar{r}_2)=\sum_{\mu>0}E_\mu(r_1,r_2)E^*_\mu(\bar{r}_1,\bar{r}_2)
g_h(y-y'),
\label{propce}
\eeq
formed by functions $E_\mu$. In complex
notations they are
\beq
E_\mu(r_1, r_2)=\Bigl(\frac{r_{12}}{r_{10}r_{02}}\Bigr)^h
\Bigl(\frac{r^*_{12}}{r^*_{10}r^*_{02}}\Bigr)^{\bar{h}},
\label{confb}
\eeq
where $\mu=\{n, \nu, r_0\}$ denotes a set of conformal weights $n$, $\nu$
and two-dimensional center-of-mass coordinate $r_0$. Integer $n$ and real $\nu$ enumerates functions of the basis.
The conformal weight defined by  ${n, \nu}$ is
\beq
h=\frac{1+n}{2}+i\nu;\,\,\,\,\,\bar{h}=1-h^*.
\eeq
For simplicity we shall also use $h$ to denote a set  $\{n,\nu\}$.
Restriction $\mu>0$ means using half of the whole set with $\nu>0$, so that
\beq
\sum_{\mu>0}=\sum_{n=-\infty}^\infty\int_0^\infty d\nu\frac{1}{a_h}
\int d^2r_0,\ \ a_h=\frac{\pi^4}{2}\frac{1}{\nu^2+n^2/4}.
\eeq

Passing from rapidity $y$ to the complex
angular momentum $j = 1+\omega$, we have the pomeron propagator
\beq
g_\omega(r_1,r_2;\bar{r}_1, \bar{r}_2)=
\int_{-\infty}^\infty dy e^{-\omega y} g_y(r_1,r_2;\bar{r}_1, \bar{r}_2)
\eeq
with the inverse transformation
\beq
g_y(r_1,r_2;\bar{r}_1,\bar{r}_2)=\int_{\sigma-i\infty}^{\sigma+i\infty}
\frac{d \omega}{2\pi i}e^{\omega y} g_\omega(r_1,r_2;\bar{r}_1,\bar{r}_2).
\eeq
In fact $-\omega$ is the "energy" corresponding to the BFKL Hamiltonian.

As a function of $\omega$,  in the conformal
basis,  the propagator is
\beq
g_{\omega,h}=\frac{1}{l_{n\nu}}\frac{1}{\omega-\omega_h},
\label{confprop}
\eeq
where $\omega_h$ are the BFKL energy levels
\beq
\omega_h=2\bar{\alpha}_s(\psi(1)-{\rm Re}\psi(h))
\eeq
and
\beq
l_h=\frac{4\pi^8}{a_{n+1,\nu}a_{n-1,\nu}}.
\eeq

The form of the triple-pomeron vertex $\Gamma$ can be extracted from the
interaction part of the Lagrangian of the effective non-local field theory
\cite{MB06}, where pomerons are described by two non-local fields
$\Phi$ and $\Phi^\dag$:
\beq
L_I=\frac{2\alpha_s^2N_c}{\pi}
\int\frac{d^2r_1d^2r_2d^2r_3}{r_{12}^2r_{23}^2r_{13}^2}
\Phi(y, r_2, r_3)\Phi(y, r_3, r_1)L_{12}\Phi^\dag(y, r_1,r_2)+h.c.,
\label{LI}
\eeq
where operator $L_{12}$ is defined as
\beq
L_{12}=r_{12}^4\nabla_1^2\nabla_2^2
\eeq
and acts on splitting pomeron or the one formed after fusion.
Passing  to the conformal representation one presents:
\beq
\Gamma(r_1,r_2|r_3,r_4;r_5,r_6)=\sum_{\mu_1,\mu_2,\mu_3>0}
\Gamma_{\mu_1|\mu_2,\mu_3}E_{\mu_1}(r_1,r_2)E^*_{\mu_2}(r_3,r_4)
E^*_{\mu_3}(r_5,r_6).
\label{confv}
\eeq
The conformal invariance of the pomeron interaction dictates the structure
of $\Gamma_{\mu_1|\mu_2,\mu_3}$ as:
\beq
\Gamma_{\mu_1|\mu_2,\mu_3}=R_{12}^{\alpha_{12}}R_{23}^{\alpha_{23}}
R_{13}^{\alpha_{13}}\times(c.c.)\times\Gamma(\bar{h}_1, h_2,h_3),
\label{confvcoef}
\eeq
where parameters $\alpha_{ik}$ are known function of conformal weights
$n$, $\nu$, see ~\cite{braun1}, and (c.c.) refers to the complex conjugate of the preceding
factors.

In the lowest order
\beq
\Gamma^{(0)}_{\mu_1|\mu_2,\mu_3}=R_{12}^{\alpha_{12}}
R_{23}^{\alpha_{23}}R_{13}^{\alpha_{13}}\times(c.c.)\times
\Omega(\bar{h}_1, h_2,h_3).
\eeq
The conformal vertex $\Omega$ was introduced and studied by Korchemsky
~\cite{KC99} and
corresponds to planar diagrams contribution, which gives the dominant part
in the limit of large number of colours. As mentioned in the Introduction,
in ~\cite{KC99} explicit formulas for $\Omega$ were derived.
In our numerical studies we use
the results of  Korchemsky in the following form (they are slightly different
than in ~\cite{KC99}):
$$
\Omega(h_1, h_2, h_3)=\pi^3\bigl[\Gamma^2(h_1)\Gamma^2(h_2)
\Gamma(1-h_1)\Gamma(1-h_2)\Gamma(1-h_3)\bigr]^{-1}
$$
\beq
\times\sum_{a=1}^3J_a(h_1,h_2,h_3)\bar{J}_a(\bar{h}_1,\bar{h}_2,\bar{h}_3),
\eeq
where functions $J_a$ and $\bar{J}_a$ are defined via
convolutions of hypergeometric functions:
$$
J_1(h_1, h_2, h_3)=\Gamma(h_1+h_2-h_3)\Gamma(1-h_1)\Gamma(h_1)
\Gamma(1-h_2)\Gamma(h_2)
$$
$$
\times\int_0^1dx\,\,(1-x)^{-h_3}{_2F_1}(h_1,1-h_1;1;x)
{_2F_1}(h_2,1-h_2;1;x);
$$
$$
J_2(h_1, h_2, h_3)=\frac{\Gamma(h_1+h_2-h_3)\Gamma(1-h_1)\Gamma(h_1)
\Gamma(1-h_2)\Gamma(h_2)\Gamma^2(1-h_3)}{\Gamma(1+h_1-h_3)\Gamma(2-h_1-h_3)}
$$
$$
\times{}_4 F_3\left(\begin{array}{llll}h_2, & 1-h_2, & 1-h_3, & 1-h_3
\\ 1, & 2-h_1-h_3, & 1+h_1-h_3\end{array}\Bigr|1\right);
$$
\beq
J_3(h_1, h_2, h_3)=J_2(h_2,h_1,h_3)
\label{vert1}
\eeq
and
$$
\bar{J}_1(\bar{h}_1, \bar{h}_2, \bar{h}_3)
=(-1)^{n_1+n_2}\frac{\Gamma(-\bar{h}_1+\bar{h}_2+\bar{h}_3)
\Gamma(1-\bar{h}_2)\Gamma(1-\bar{h}_1)\Gamma(\bar{h}_1)}
{\Gamma(-\bar{h}_1+\bar{h}_3+1)}
$$
$$
\times\int_0^1dx\,\,x^{-\bar{h}_1}(1-x)^{\bar{h}_2-1}{_2F_1}
(\bar{h}_2,-\bar{h}_1+\bar{h}_2+\bar{h}_3;-\bar{h}_1+\bar{h}_3+1;x){}_2F_1
(1-\bar{h}_3, 1-\bar{h}_1;1;x);
$$
$$
\bar{J}_2(\bar{h}_1, \bar{h}_2, \bar{h}_3)=(-1)^{n_1}
\frac{\Gamma(1-\bar{h}_1)\Gamma(\bar{h}_3)
\Gamma(-\bar{h}_1+\bar{h}_2+\bar{h}_3)
\Gamma(1-\bar{h}_2)}{\Gamma^2(-\bar{h}_1+\bar{h}_3+1)}
$$
$$
\times\int_0^1dx\,\,x^{\bar{h}_3-\bar{h}_1}(1-x)^{\bar{h}_2-1}
{_2F_1}(1-\bar{h}_1,1-\bar{h}_1;-\bar{h}_1+\bar{h}_3+1;x){}_2F_1
(\bar{h}_2, -\bar{h}_1+\bar{h}_2+\bar{h}_3;-\bar{h}_1+\bar{h}_3+1;x);
$$

\beq
\bar{J}_3(\bar{h}_1, \bar{h}_2, \bar{h}_3) =
\bar{J}_3(\bar{h}_2, \bar{h}_1, \bar{h}_3).
\label{vert2}
\eeq

In any pomeron Feynman diagram BFKL propagators and triple vertices
are convoluted by integration over  coordinates
or momenta.
One can perform this integration using expansions (\ref{propce}),
(\ref{confv}) and completeness condition for the conformal basic functions:
\beq
\int\frac{d^2r_1d^2r_2}{r^4_{12}}
E_\mu(r_1,r_2) E^*_{\mu'}(r_1,r_2) =\delta_{\mu\mu'}.
\eeq
As a result this convolution is substituted by the summation over conformal
weights and integration over intermediate center-of-mass coordinates.
The latter can be done analytically using the known dependence
(\ref{confvcoef}).

In the energetic representation
all propagators become  functions of $\omega$.
The interaction vertex conserves the energy.
Indeed, a product of two pomeron propagators, which form the simplest loop,
contains the following integration over intermediate energy in
$\omega$-space (note that integration over gluon coordinates is suppressed):
\beq
\int dy e^{-\omega y}g_1(y)g_2(y)=
\int_{\sigma-i\infty}^{\sigma+i\infty}
\frac{d\omega_1}{2\pi i}g_1(\omega_1)g_2(\omega-\omega_1).
\eeq
It is obvious that number of integrals over energy corresponds to the
number of loops of the Feynman diagram.

With a help of this rules one can immediately write down the Schwinger-Dyson
equation for a full conformal pomeron propagator $G_{\omega,h}$, which
includes arbitrary number of  loop insertions:
\beq
G_{\omega,h}=g_{\omega,h}-g_{\omega,h}l^2_h\Sigma_{\omega,h}G_{\omega,h},
\label{fconfprop}
\eeq
where $l_h^2$ comes from the triple-pomeron vertex operator $L$ acting
on the incoming and outgoing pomeron propagators.
Function $\Sigma_{\omega, h}$ is the pomeron self-mass in the conformal basis:
\beq
\Sigma_{\omega\mu}=\frac{8\alpha_s^4N_c^2}{\pi^2}
\int\frac{d\omega_1}{2\pi i}\sum_{\mu_1\mu_2}
\Gamma^{(0)}_{\mu|\mu_1,\mu_2}G_{\mu_1}G_{\mu_2}
\Gamma_{\mu_1,\mu_2|\mu}.
\eeq
Using (\ref{confvcoef}) one can perform integrations over intermediate
center-of-mass coordinate explicitly. As a result we are left with
\beq
\Sigma_{\omega,h}=\frac{8\alpha_s^4N_c^2}{\pi^2}
\int\frac{d\omega_1}{2\pi i}\sum_{h_1h_2}
\Gamma^{(0)}_{h|h_1,h_2}G_{\omega,h_1}G_{\omega,h_2}\Gamma_{h_1,h_2|h},
\label{sigma}
\eeq
where
\beq
\sum_h=\frac{2}{\pi^4}\sum_{n=-\infty}^{\infty}
\int_0^{\infty}d\nu\Bigl(\nu^2+\frac{n^2}{4}\Bigr).
\eeq
The formal solution of equation (\ref{fconfprop}) is
\beq
G_{\omega,h}=\frac{1}{1/g_{\omega,h}+l_h^2\Sigma_{\omega,h}}.
\label{sdsol}
\eeq
In the lowest order one should change $G_{\omega,h}\to g_{\omega_h}$
and $\Gamma_{h_1h_2|h}\to\Gamma^{(0)}_{h_1h_2|h}$ in (\ref{sigma}).

\section{Scattering amplitude}

The  amplitude for the scattering of two hadrons at rapidity $y$
can be presented as
\beq
A(s, t)=is\int_{\sigma-i\infty}^{\sigma+i\infty}\frac{d\omega}{2\pi i}
s^\omega f_\omega(q^2),
\eeq
where function $f_\omega(q^2)$ is a $t$-channel partial wave for fixed
transferred momentum.

In the approximation of a single bare pomeron exchange it is a convolution
of the bare pomeron propagator $g_\omega$ with two impact factors
describing
interaction of the pomeron with external hadrons.
Passing to the coordinate space we have
\beq
(2\pi)^2f_\omega(q^2)\delta^{(2)}(q-q')=
\int d^2r_1 d^2r_2 d^2r_3 d^2r_4\,\,\, \Phi_1(r_1, r_2, q)
g_\omega(r_1,r_2|r_3,r_4)\Phi^*_2(r_3,r_4,q').
\label{amp}
\eeq

The impact factors $\Phi_i(r_1,r_2, q)$ in the coordinate space
 in (\ref{amp}) are related to those in the momentum space by
\beq
\Phi(r_1,r_2,q)=e^{iqR}\Phi(r,q),\ \ \Phi(r,q)=\int\frac{d^2 k}{(2\pi)^2}\Phi(k, q-k)
e^{i(k-\frac{q}{2})r},
\label{fimp}
\eeq
where $R=(r_1+r_2)/2$ and $r=r_{12}$.
Note that  for a colorless object the impact factor $\Phi(k, q-k)$ in the
momentum space have to vanish at $k=0$. Correspondingly in the
coordinate space integration of $\Phi(r,q)$ over $r$ has to give zero.
This can always be achieved by changing
\beq
\Phi(r,q)\to \Phi(r,q)-\delta^2(r)\int d^2r\Phi(r,q).
\label{fimpnew}
\eeq
However the term with the $\delta$-function does not contribute, since
the pomeron propagator in (\ref{amp}) vanishes at $r_{12}=0$ or $r_{34}=0$.
So we are not to bother about this property of $\Phi(r,q)$ for a colorless
target provided it is integrable over $r$.

Putting (\ref{fimp}) into (\ref{amp}) and separating the
dependence on center-of-mass coordinates we get
\beq
f_\omega(q^2)=\int d^2r d^2r'\,\,\,
\Phi_1(r, q) g^q_\omega(r,r') \Phi^*_2(r', q),
\eeq
where the pomeron propagator in the mixed representation is
\beq
(2\pi)^2g^q_\omega(r,r')\delta^{(2)}(q-q')
=\int d^2R d^2R'\,\, e^{iqR}e^{-iq'R'}g_\omega(r_1,r_2|r_3,r_4).
\eeq
Here $g_\omega^q(r,r')$ describes propagation of the pomeron
with momentum transfer $q$ and can be interpreted as the amplitude for
the scattering of two dipoles with sizes $r$ and $r'$.

The expression $g_\omega^q(r,r')$ in the conformal basis
can be easily found if one uses the corresponding conformal expansion for the
bare propagator. Indeed the pomeron Green function in the conformal basis is
\beq
g_\omega(r_1,r_2|r_3,r_4)=\frac{2}{\pi^4}
\sum_n\int d^2r_0\int_0^\infty d\nu \Bigl(\nu^2 +
\frac{n^2}{4}\Bigr)g_{\omega,h} E_\mu(r_{10}, r_{20})
E^*_\mu(r_{30}, r_{40}),
\eeq
so that
$$
g^q_\omega(r,r')=\frac{2}{\pi^4}\sum_n\int_0^\infty d\nu
\Bigl(\nu^2 + \frac{n^2}{4}\Bigr)g_{\omega,h}
$$
\beq
\times \int d^2R\,\, e^{iqR}E_\mu(R+\frac{r}{2}, R
-\frac{r}{2})\int d^2R'\,\, e^{-iqR'}E^*_\mu(R'+\frac{r'}{2},
R'-\frac{r'}{2}).
\label{fourprop}
\eeq

For our purposes it will be sufficient co consider scattering at zero
transferred momentum, i.e. forward scattering amplitude $f_\omega(0)$.
For this we have to know behavior of the last two integrals in
(\ref{fourprop}) at $q^2\to 0$. The Fourier transformation of conformal
functions (\ref{confb}) in the region of small momentum transfer was
investigated in \cite{LL86}. One can use this results to pass to the limit
$q=0$:
\beq
g^0_\omega(r,r')=\frac{1}{\pi^2}|rr'|\sum_n\int_0^\infty d\nu
\Bigl|\frac{r}{r'}\Bigr|^{2i\nu}\Bigl(\frac{r^*r'}{rr'^*}
\Bigr)^{n/2}g_{\omega,h}.
\eeq
We shall be interested only in the leading contribution, which comes from
$n=0$. With $n=0$
\beq
g^0_\omega(r,r')= \frac{1}{\pi^2}\int_0^\infty d\nu \,\,
|r|^{1+2i\nu}|r'|^{1-2i\nu}g_{\omega,\nu}.
\eeq
Therefore the forward scattering amplitude is
\beq
f_\omega(0)=\frac{1}{\pi^2}\int_0^\infty d\nu \,\,
g_{\omega,\nu} \int d^2r \Phi(r, 0) |r|^{1+2i\nu} \int d^2r'
\Phi^*(r', 0) |r'|^{1-2i\nu}.
\label{forwamp}
\eeq

It is natural to choose the impact factor in a Gaussian form
\beq
\Phi(r, 0)=\frac{\lambda b}{\pi}e^{-b r^2},
\label{impchoice}
\eeq
where $b$ is the inverse of the hadron radius squared $b = 1/R^2_N$ and
dimensionless $\lambda$ is the effective coupling of the hadron to pomeron.
The value of $\lambda$ is obviously irrelevant for the study of the
relative contribution of loops.
With  choice (\ref{impchoice}) integration over $r$ can be trivially done:
\beq
\int d^2r\Phi(r, 0)|r|^{1+2i\nu}=\frac{\lambda b}{\pi}
\pi b^{-3/2-i\nu}\Gamma(3/2+i\nu).
\eeq
Putting this result into (\ref{forwamp}) we find
\beq
f_\omega(0)=\frac{1}{\pi^2}\int_0^\infty d\nu \,\,
g_{\omega,\nu}\times \frac{\lambda^2}{b}
\Bigl(\nu^2 + \frac{1}{4}\Bigr)\frac{\pi}{\cosh(\pi\nu)},
\eeq
where $g_{\omega,\nu}$ is a conformal propagator (\ref{confprop})
with $n=0$.

To take into account the loop contribution one has to substitute the
bare propagator $g_{\omega,\nu}$ by the full Green function. With a single loop
insertion
\beq
G_{\omega,\nu}=\frac{1}{1/g_{\omega,\nu}+l_{0\nu}^2
\Sigma_{\omega,\nu}}=
\frac{1}{l_{0\nu}}\frac{1}
{\omega-\omega_\nu}-\frac{\Sigma_{\omega,\nu}}
{(\omega-\omega_\nu)^2},
\label{sdexp}
\eeq
where now
\beq
\omega_\nu=2\bar{\alpha}_s(\psi(1)-{\rm Re}\,\psi(1/2+i\nu))
\eeq
and
\beq
\frac{1}{l_{0\nu}}=\frac{1}{16}\frac{1}{(\nu^2+1/4)^2}.
\eeq

The first term in (\ref{sdexp}) comes from exchange of the bare pomeron.
It is not difficult to show that in terms of rapidity $y$ it gives
\beq
A^{(1)}_y(0)=\frac{1}{16\pi^2}
\frac{\lambda^2}{b}\int_0^\infty d\nu \,\,
\frac{1}{\Bigl(\nu^2 + \frac{1}{4}\Bigr)}
\frac{\pi}{\cosh(\pi\nu)}e^{\omega_\nu y}.
\label{f1amp}
\eeq
The second term in (\ref{sdexp}) corresponds to the lowest order loop
contribution. As a function of $\omega$
\beq
f^{(2)}_\omega(0) =-\frac{1}{16\pi^2}\frac{\lambda^2}{b}
\int_0^\infty d\nu\,\,16\Bigl(\nu^2+\frac{1}{4}\Bigr)
\frac{\Sigma_{\omega,\nu}}{(\omega-\omega_\nu)^2}\frac{\pi}{\cosh(\pi\nu)}
\eeq
with the explicit form of $\Sigma_{\omega,\nu}$ given in
\cite{MB09}:
\beq
\Sigma_{\omega,\nu}=\frac{\alpha^4_sN^2_c}{8\pi^{10}}
\int_0^\infty d\nu_1 d\nu_2 \frac{\nu_1^2}{\Bigl(\nu_1^2 +
\frac{1}{4}\Bigr)^2}
\frac{\nu_2^2}{\Bigl(\nu_2^2 + \frac{1}{4}\Bigr)^2}
\frac{\Omega^2(1/2+i\nu,1/2+i\nu_1,1/2+i\nu_2)}
{\omega-\omega(0,\nu_1)-\omega(0,\nu_2)}
\eeq
and the conformal vertex $\Omega$ given by (\ref{vert1}) and (\ref{vert2}).

For numerical calculations it is convenient to pass to rapidity. Performing
integration over $\omega$  we find
$$
A^{(2)}_y(0)=-\frac{1}{16\pi^2} \frac{\lambda^2}{b}
\int_{0}^\infty d\nu \,\, 16\Bigl(\nu^2 +
\frac{1}{4}\Bigr) \frac{\pi}{\cosh(\pi\nu)}
$$
$$
\times \frac{\alpha^4_sN^2_c}{8\pi^{10}}\int_0^\infty d\nu_1 d\nu_2
\frac{\nu_1^2}{\Bigl(\nu_1^2 + \frac{1}{4}\Bigr)^2}\frac{\nu_2^2}
{\Bigl(\nu_2^2+\frac{1}{4}\Bigr)^2}\Omega^2(1/2+i\nu,1/2+i\nu_1,1/2+i\nu_2)
$$
\beq
\times\Bigl(\frac{e^{\omega_\nu y}y}{\omega_\nu-\omega_{\nu_1}-
\omega_{\nu_2}}-\frac{e^{\omega_{\nu_1} y}}
{\bigl(\omega_\nu-\omega_{\nu_1}-\omega_{\nu_2}\bigr)^2}+
\frac{e^{(\omega_{\nu_1}+\omega_{\nu_2}) y}}
{\bigl(\omega_\nu-\omega_{\nu_1}-\omega_{\nu_2}\bigr)^2}\Bigr).
\label{f2amp}
\eeq
The total forward scattering amplitude with the lowest order loop
correction is a sum of (\ref{f1amp}) and (\ref{f2amp}):
\beq
A_y(0)=A^{(1)}_y(0)+A^{(2)}_y(0).
\eeq
Numerical calculations of (\ref{f1amp}) and (\ref{f2amp}) will be presented in the next section.

\section{Numerical studies}

We have set up a program which calculates the
 bare pomeron exchange amplitude (\ref{f1amp}) and single-loop contribution
(\ref{f2amp}).
By far the most difficult part is the computation of the triple-pomeron
vertex $\Omega$. It requires complicated numerical procedures and is extremely
time-consuming. The vertex is a complex function which depends on
three  conformal variables $\nu$. We have restricted this variables
to lie in the interval
$0<\nu<3.0$ and introduced a grid dividing this interval into $N$ points.
We found the vertex on the grid, which requires calculation at $N^3$ points.
This strongly limits  numbers $N$ admissible for given
calculation resources. In our case we used $N=30$ compatible with
reasonable computation time. The value of the vertex in between the grid
points was found by  interpolation.

The integrals in (\ref{f1amp}) and (\ref{f2amp}) were calculated by the
Newton-Cotes integration formulas. The limits in $\nu$ were taken as
before $0<\nu<3.0$. The number of sample points was chosen to provide
relative error of the order of $10^{-3}$. The same integration strategy was used for the
vertex integrals in (\ref{vert1}), (\ref{vert2}).

\begin{figure}
\hspace*{2 cm}
\epsfig{file=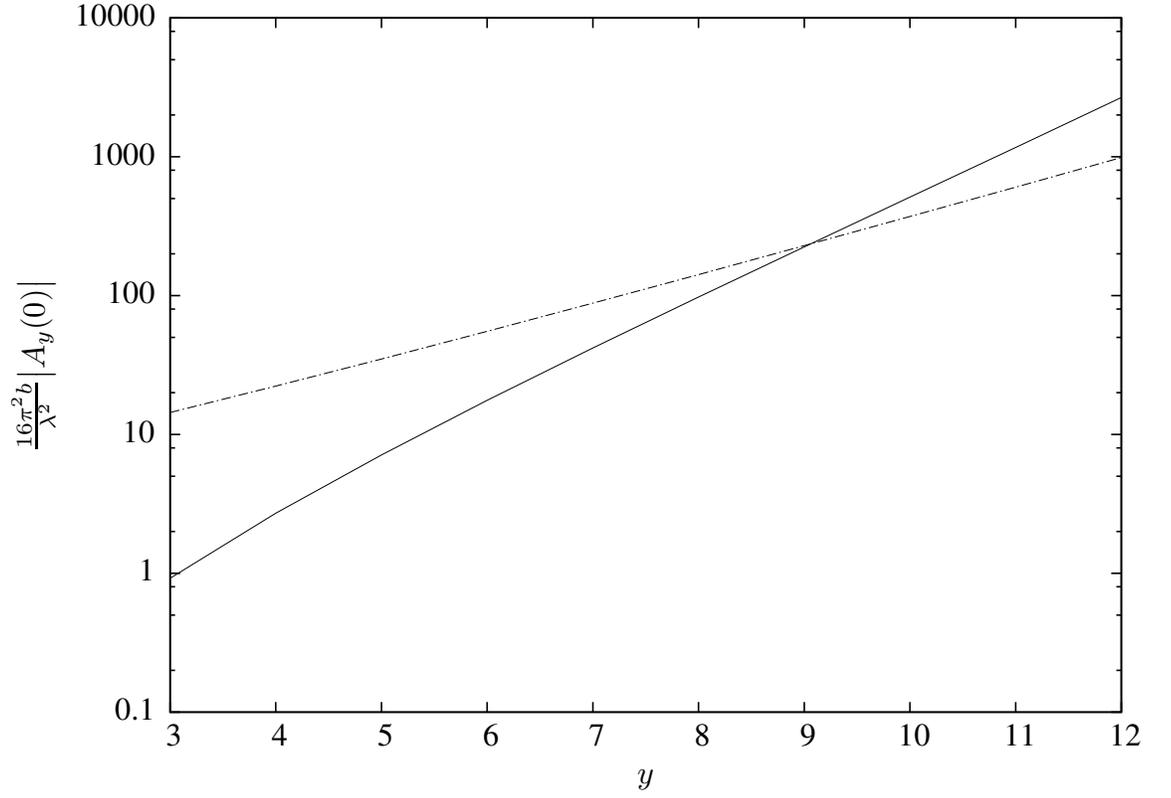,width=15 cm}
\caption{The bare pomeron exchange (\ref{f1amp}) (dash-dotted curve)
and  single-loop (\ref{f2amp}) (solid curve) contributions to the
forward scattering amplitude as functions of rapidity.}
\label{fig1}
\end{figure}

We have performed calculations for the standard value of the QCD coupling
constant $\alpha_s = 0.2$ and $N_c = 3$. In Fig. \ref{fig1} the dash-dotted
line presents the bare pomeron exchange contribution
(\ref{f1amp}). The behavior of the  amplitude is determined by the
initial pole of the conformal BFKL propagator (\ref{confprop}).
As expected, the curve grows with rapidity roughly as $e^{\Delta y}$, where
$\Delta\approx 0.48$. The behavior of the single-loop contribution
(\ref{f2amp}) is shown in Fig.  \ref{fig1} by the
solid line. It roughly grows twice
faster, as $\sim e^{2\Delta y}$, again as expected. Note that the
results in Fig.  \ref{fig1} are normalized by
 factor $16\pi^2\,b/\lambda^2$.

\begin{figure}
\hspace*{2 cm}
\epsfig{file=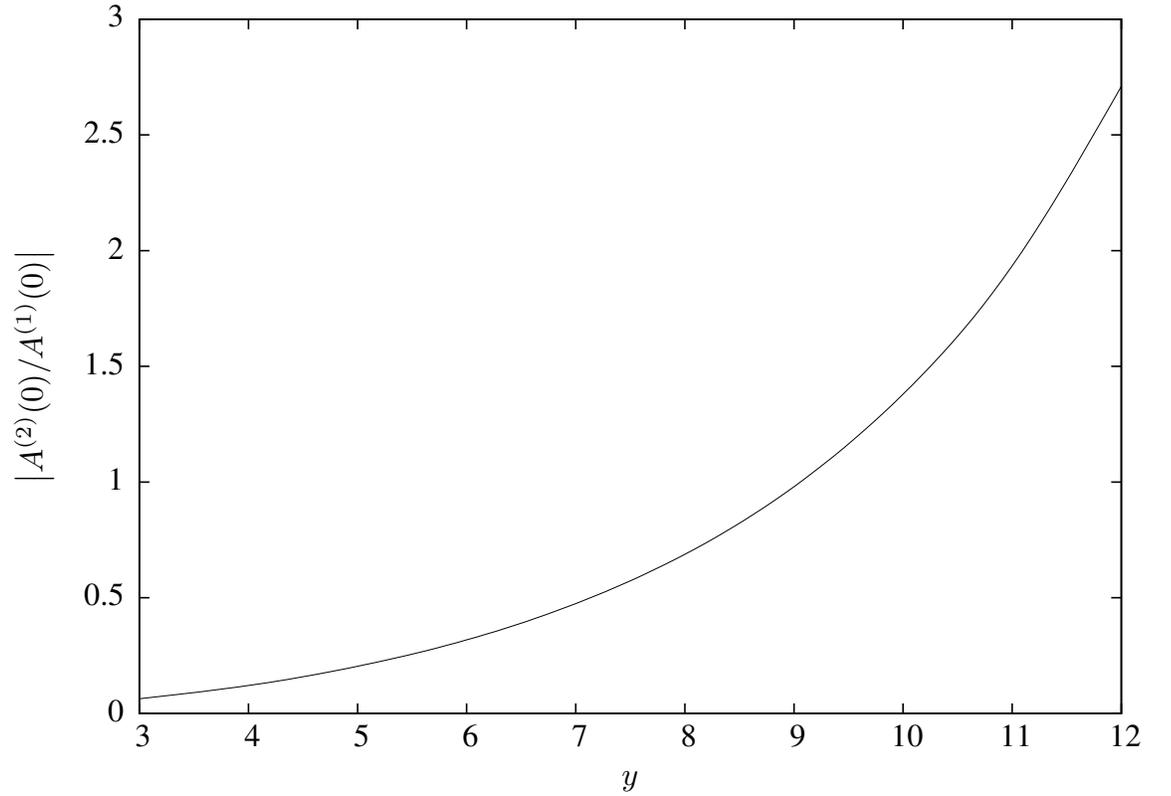,width=15 cm}
\caption{The ratio $|A^{(2)}_y(0)/A^{(1)}_y(0)|$ as a function of rapidity.}
\label{fig2}
\end{figure}

For small rapidities the loop term is suppressed by the smallness of the
QCD coupling constant. However, its faster growth with rapidity
compensates this very early. This is illustrated in
Fig.  \ref{fig2}, where  the ratio
$R(y)=|A^{(2)}_y(0)/A^{(1)}_y(0)|$ is shown (recall that
this
quantity does not depend on the hadron radius nor effective coupling of the
hadron to pomeron $\lambda$). One clearly  observes the exponential
growth of R:
$R(y)\sim \exp (\Delta y)$.

As a result we  conclude that the loop contribution becomes
visible already at rapidities $3-8$ and starts to dominate at $y\sim8-10$.

\begin{figure}
\hspace*{2 cm}
\epsfig{file=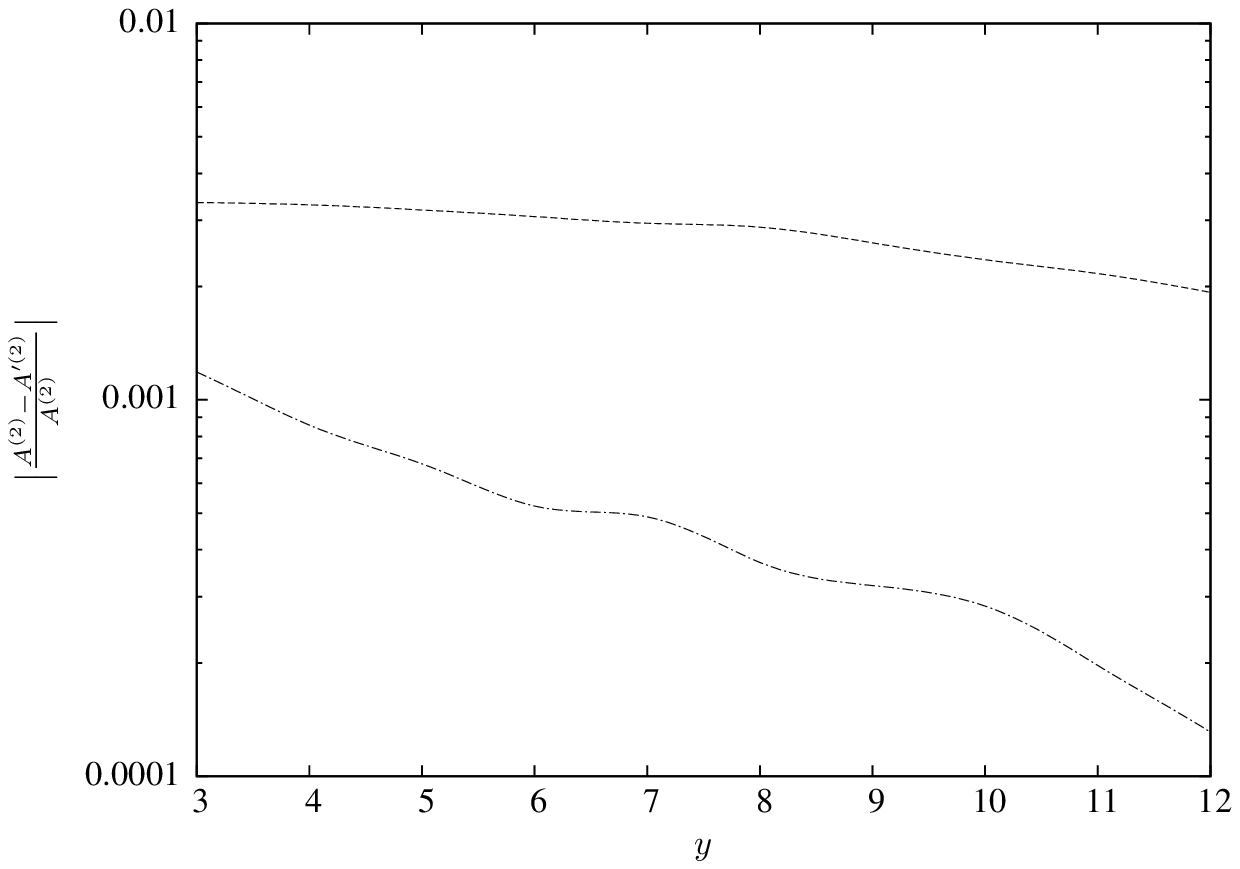,width=15 cm}
\caption{The ratio $|(A^{(2)}-A'^{(2)})/A^{(2)}|$ as a function of rapidity, where $A^{(2)}$ and $A'^{(2)}$ are single-loop contributions (\ref{f2amp}) calculated at different $N$ and maximum $\nu$. For both curves $A^{(2)}$ corresponds to our basic set $0<\nu<3.0$, $N=30$ and $A'^{(2)}$ was found at $0<\nu<2.0$, $N=20$ and $0<\nu<3.0$, $N=20$ for dash-dotted and dashed lines correspondingly.}
\label{fig3}
\end{figure}

To illustrate the convergence of our results with respect to the value of $N$ and maximum $\nu$, in Fig. \ref{fig3} we show a ratio $|(A^{(2)}-A'^{(2)})/A^{(2)}|$ of single-loop contributions $A^{(2)}$ and $A'^{(2)}$ (\ref{f2amp}) found at different values of this parameters. In case of both curves for $A^{(2)}$ we use a basic set  $0<\nu<3.0$, $N=30$ from Figs. \ref{fig1} and \ref{fig2}. To show sensitivity of our results to the choice of $\nu_{max}$, we calculate $A'^{(2)}$ using a different interval $0<\nu<2.0$ and $N=20$ to keep a size of the grid cell used to calculate $\Omega$ unchanged. The result is presented by the dash-dotted line. It is easy to see that an error caused by the choice of $\nu_{max}$ lies within the accuracy of calculation of integrals in (\ref{f2amp}) (relative error of the order of $10^{-3}$). One can come to the same conclusion from analysis of the dashed line, which presents dependence of our results on the choice of the size of the grid ($N$). In this case for calculation of $A'^{(2)}$ we use a larger cell size, i.e. $0<\nu<3.0$, $N=20$.

\section{Conclusions}

We have studied a single loop contribution to the scattering amplitude of
two colliding hadrons. We have found expression for the amplitude in a
framework of conformal invariant technique with more or less general form
of the impact factors. The triple-pomeron vertex with full dependence on the
intermediate conformal weights was calculated and used.

Numerical analysis shows that smallness of the QCD coupling constant
is compensated by rapid growth of the single-loop amplitude with rapidity.
We found that the loop contribution manifests itself at relatively small
rapidities $y\sim3-8$ and dominates the bare pomeron exchange amplitude
already at $y\sim8-10$. In general this conclusions are in accordance
with \cite{MB09}. Note that a similar conclusion was reached in the framework
of one-dimensional model for QCD in ~\cite{iancu}, where it was found that
pomeron loops at fixed coupling  become of relative order one when $Y\sim 5$.

Thus loops have to be taken into account already at present
energies. Higher order loops calculation is required for larger values of
rapidity with summation of all loops in the limit $y\to\infty$.
As mentioned in the Introduction this nontrivial problem is left for future investigations
together with inclusion of other type of scattering hadron and running coupling.

\section{Acknowledgements}
This work has been supported by the RFFI grant 12-02-00356-a
and the SPbSU grants 11.059.2010, 11.38.31.2011 and 11.38.660.2013.

\end{document}